\documentclass[global,twocolumn]{svjour-1}
\usepackage{graphicx}
\usepackage{color}
\usepackage{colortbl}

\newcommand{\un}[1]{\:\mathrm{#1}}
\journalname{Applied Physics A: Materials Science and Processing}
\begin{document}
\title{
 {\vskip -35mm
        {\small  \textcolor{blue}{10th International Conference on Laser Ablation 22-27 Nov. 2009, Singapore,
                     http://cola2009.org \\
                     Invited report submitted to Applied Physics A: Materials Science and Processing (2010) }
                     }\\
  \vskip -5mm
  \rule{170mm}{0.2mm}
  \vskip  2mm
  }
           Spallative ablation of dielectrics by X-ray laser}
%\subtitle{Do you have a subtitle?\\ If so, write it here}
\author{N.~A.~Inogamov\inst{1}\thanks{\emph{email:} nailinogamov@googlemail.com (N.~Inogamov)},
        V.~V.~Zhakhovsky\inst{2,3},
        A.~Ya.~Faenov\inst{4,3},
        V.~A.~Khokhlov\inst{1},
        V.~V.~Shepelev\inst{5},
        I.~Yu.~Skobelev\inst{3},
        Y.~Kato\inst{4,6},
        M.~Tanaka\inst{4},
        T.~A.~Pikuz\inst{4,3},
        M.~Kishimoto\inst{4},
        M.~Ishino\inst{4},
        M.~Nishikino\inst{4},
        Y.~Fukuda\inst{4},
        S.~V.~Bulanov\inst{4},
        T.~Kawachi\inst{4},
        Yu.~V.~Petrov\inst{1},
        S.~I.~Anisimov\inst{1},
        V.~E.~Fortov\inst{3}
% \thanks is optional - remove next line if not needed
}                     % Do not remove
\institute{ L.D. Landau Institute for Theoretical Physics, Russian Academy of Sciences, Chernogolovka 142432,
Russia \and
            Department of Physics, University of South Florida, Tampa, Florida 33620-5700, USA \and
            Joint Institute for High Temperatures, Russian Academy of Sciences, Moscow 125412, Russia \and
            Kansai Photon Science Institute, Japan Atomic Energy Agency, Kyoto 619-0215, Japan \and
            Institute for Computer Aided Design, Russian Academy of Sciences, Moscow 123056, Russia \and
            The Graduate School for the Creation of New Photonics Industries, Hamamatsu, Shizuoka 431-1202,
             Japan }
\authorrunning{Inogamov, Zhakhovsky, Faenov et al.}
 \titlerunning{Spallative ablation}
\date{Received: 27-Nov-2009 / Revised version: date}
% The correct dates will be entered by the editor
%
\maketitle
\begin{abstract}
 Short laser pulse in wide range of wavelengths, from infrared to X-ray,
   disturbs electron-ion equilibrium and rises pressure in a heated layer.
 The case where pulse duration $\tau_L$ is shorter than acoustic relaxation time $t_s$
   is considered in the paper.
 It is shown
   that this short pulse may cause thermomechanical phenomena such as spallative ablation
   {\it regardless} to wavelength.
 While the physics of electron-ion relaxation {\it strongly depends} on wavelength
     and various electron spectra of substances:
       there are spectra with an energy gap in semiconductors and dielectrics
         opposed to gapless continuous spectra in metals.
 The paper describes entire sequence of thermomechanical processes
   from expansion, nucleation, foaming, and nanostructuring to spallation
     with particular attention to spallation by X-ray pulse.

\end{abstract}

%
%%%%%%%%%%%%%%%%%%%%%%%%%%%%%%%%%%%%%%%%%%%%%%%%%%%%%%%%%%%%%%%%%%%%%%%%%%%%%%%%%%%%%%%%
%
\section{Introduction}
\label{intro}

 % a-01

 There are many industrial applications using short pulse lasers.
 New exciting possibilities are connected with development of the X-ray lasers.
 Ablations by the long and short pulses differ qualitatively.
 The first - evaporate, boil, and, at higher fluences, move matter by ablative pressure
   created in hot plasma corona.
 While the release of pressurized layer is the main process in the case of short pulse.
 It is shown below that this is true for any laser wavelength.

 % a-02
 Let's consider short pulse.
 Irradiation with sufficient intensity transfers substance into warm dense matter state.
 In condensed phase, the cohesive properties are important.
 The cohesion is result of interatomic attraction.
 Due to stiff behavior of solids and liquids, their expansion in rarefaction wave
   is very different from expansion of gas.
 Stiff means that the typical moderate expansions $\Delta\rho/\rho\sim (0.1-0.2)$
   cause the order of magnitude pressure drops
     and even change of sign of pressure - from compressed to stretched state,
       here $\Delta\rho$ is the density drop due to expansion.
 Stretched metastable layer under tensile stress appears as a result of the release of the warm dense matter
   with its stiff response.

 % a-03   c=1

 In metastable state substances are sensitive to temperature and degree of stretching.
 Nucleation probability exponentially depends on amplitude of negative pressure.
 Therefore, sharp nucleation threshold appears on this amplitude.
 The threshold depends on temperature.
 Nucleation is followed by development of two-phase layer
   composed of condensed phase and voids.
 If the laser heated layer melts, then nucleation takes place in liquid.
 In this case, expansion of the two-phase layer leads to foaming.
 The foaming may cause formation of the surface nanorelief \cite{nanoreliefLETTjetp08}.

 % a-04
 The metastability and nucleation are
   % usual
   % familiar
   well-known for release initiated by shock coming to the surface from the bulk of a target.
 But in the case of shock
   sent by a long laser pulse, an ion beam, or by an explosion of chemical explosive,
     near nucleation threshold
       the two-phase layer
         locates far from the target surface.
 In the case of large scale,
   the development of the two-phase layer cannot disturb the surface of a target
     because the nucleation layer and the target surface
       are far from each other and are independent from each other.
 On the contrary, the short pulse lasers initiate foaming very close to the surface -
   since the attenuation depth $d_{att}$ for X-ray photons,
     or thickness of a skin layer $\delta_{skin},$
       may be as small as ten nanometers.
 In this case the foaming strongly interacts with the surface,
   eventually producing frozen surface structures.
 Therefore we can use such terms as nanofoam, nanostructures, or nanospallation
   to describe the situation with small depth $d_{att}.$

 % a-05 diel 2 3 4 5

 Photon absorption and collisional processes are defined by photon energy and electronic structure.
 Infrared (IR) and visible radiation excite valent electrons
   whereas X-rays are absorbed mainly by internal shells in the one-photon interactions.
 This is why the X-ray absorption is qualitatively similar for metals, semiconductors and dielectrics.
 Action of quanta from the one electron-Volt range of energy
   depends strongly on existence of the forbidden gap $\Delta.$
 In metals, they are absorbed mainly via inverse Bremsstrahlung in the skin layer.
 In cases with gap, the seed Keldysh ionization, inverse Bremsstrahlung heating of ionized electrons
   and electron avalanche control the rise of number of conduction electrons $n_e(x,t)$
     \cite{diel-breakd-vdLinde,diel-breakd-Rubenchik,diel-breakd-Quere,diel-breakd-Rethfeld}.
 Significant heating takes place when plasma frequency for these electrons $\omega_{pl}(n_e)$
   overcomes laser frequency $\omega_L$
     during the laser pulse.
 During the rest of the pulse,
   the substance with a gap absorbs laser energy
     in a skin layer.
 This means that
   during the rest of the pulse
     the absorption becomes similar to the absorption of metals.
 This greatly increases spatial density J$\cdot$cm$^{-3}$ of absorbed energy.
 Estimates show that,
   in these conditions,
     laser electric field strength
       is comparable to atomic fields of external electrons,
         their wave functions are distorted by electromagnetic wave,
           and probabilities of the multiphoton and tunnel ionizations are significant.

 % a-06

 Relaxation of electrons to equilibrium,
   after the end of a pulse,
     depends on a band structure.
 In metals, free electrons cool due to electron-ion energy transfer
   and due to thermal conductivity.
 The same is true for semiconductors which pass to metallic state during their melting.
 In substances
   which keep the gap after a pulse,
     the concentration of free electrons $n_e$ and their temperature $T_e$
       decrease after the end of a heating laser pulse
         as a result of recombination, diffusion of electrons and holes,
           and electron heat conduction.
 Three-body recombination is usually more significant than radiative recombination.
 Relaxation time $t_{eq}$ is of the order of 1-10 ps for all these cases.

 % a-07

 An acoustic response time $t_s$
   is necessary to decrease pressure few times
     in the laser-heated layer
       with thickness $d_T.$
 It equals to $t_s=d_T/c_s$ in bulk targets,
   or $t_s=d_f/c_s$ in foils with small thickness $d_f<d_T,$
     where $c_s$ is sound velocity.

 % a-08

 Electrons are light, their velocities are high
   and the electronic thermalization time $\tau_e$ is small -
     typically it is at the femtosecond range.
 In metals, this corresponds to rather high values of $T_e\sim 1$eV.
 This case is considered here.
 At room temperatures $T_e\sim 300$K the e-e relaxation in metals is slow,
       since e-e collision frequency $\nu_{ee}$ is small.
 If $\tau_e\sim 1-10$fs
   then at the picosecond time scale we have two thermodynamic subsystems: electrons and ions.
 Total pressure $p=p_e+p_i$ is composed of partial pressures.
 The contribution $p_e>0$ because $T_e-T_i>0;$ in the one-temperature state we have $p_e=0.$

 % a-09

 In our conditions,
   typical absorbed energy is $E_{abs}\sim (0.1-1) E_{coh};$
     where $E_{coh}$ is heat of sublimation; e.g., for Al $E_{coh}\approx 3$eV/atom.
 Then,
   after e-i relaxation,
     ion temperatures are in the kiloKelvin (kK) range:
   $$
   T_i\approx 4 \,(E_{abs}/\,1{\rm eV}\cdot{\rm atom}^{-1}) \, {\rm kK,}
   $$
\begin{equation}
   p_i\approx 19 \,(E_{abs}/\,1{\rm eV}\cdot{\rm atom}^{-1}) (n/6\times 10^{22}{\rm cm}^{-3})\,{\rm GPa,}
             \label{eq:pi}
\end{equation}
    if heat capacity is $\approx 3 k_B,$ and Gruneisen parameter is $\Gamma_i\sim 2.$
    The parameter $\Gamma=V(\partial p/\partial E)_V$
      links pressure rise and fast absorption of energy at the isochoric stage.

 % a-10

 At the two-temperature stage $t<t_{eq},$
   electron temperatures $T_e$ are much higher than $T_i.$
 For metals,
   in the Fermi-gas approximation,
     we have $T_e\sim\sqrt{2 E_e/\gamma}, $
   $$
   T_e=17 \,Z^{-1/6} (n/6\times 10^{22}{\rm cm}^{-3})^{1/3}(E_e/\,1{\rm eV}\cdot{\rm atom}^{-1})^{1/2}{\rm kK}
   $$
    for $t_e=T_e/T_F<1,$ $k_B T_F=E_F.$
 Here $E_e\sim E_{abs}$ is electron energy per atom, not per electron.
 Z is the number of electrons per ion,
   $\gamma=\pi^2 n_e k_{B}^2/2 E_F$
     is the electron heat capacity constant
       written in the Fermi-gas approximation.

 % a-11

 Electronic contribution $p_e$ is significant when $T_e\gg T_i,$
   then $T_e-T_i\approx T_e.$
 In this case, in metals,
\begin{equation}
       \frac{p_e}{p_{au}} = \frac{2}{5} Z \frac{n}{n_{au}} \frac{E_F}{E_{au}}
       \left(\sqrt{1+t_e^2 \frac{5\pi^2/6+25 t_e^2/4}{1+t_e^2}}-1\right),
          \label{eq:pe}
\end{equation}
 where $t_e$ is normalized temperature $T_e,$
   and $p_{au}=2.94\times 10^4$GPa,
      $n_{au}=6.757\times 10^{24}$cm$^{-3},$
        $1/n_{au}^{1/3}=0.53$\AA (Bohr radius),
          $E_{au}=27.2$eV are atomic units (au) for pressure, density, etc,
 $$
 E_F/E_{au}=(1/2)(3\pi^2 Z n/n_{au})^{2/3}
 $$
  is Fermi energy.
 The interpolation (\ref{eq:pe}) has the right limits for $t_e\ll 1,$ $t_e\gg 1.$
 \begin{equation}
 p_e\approx 2.8\, Z^{1/3} (n/6\times 10^{22}{\rm cm}^{-3})^{1/3} (T_e/1{\rm eV})^2 \, {\rm GPa}
           \label{eq:pe-2}
\end{equation}
 if $t_e<1.$

\begin{figure}[th]
 \centering
  \includegraphics[width=1.0\columnwidth]{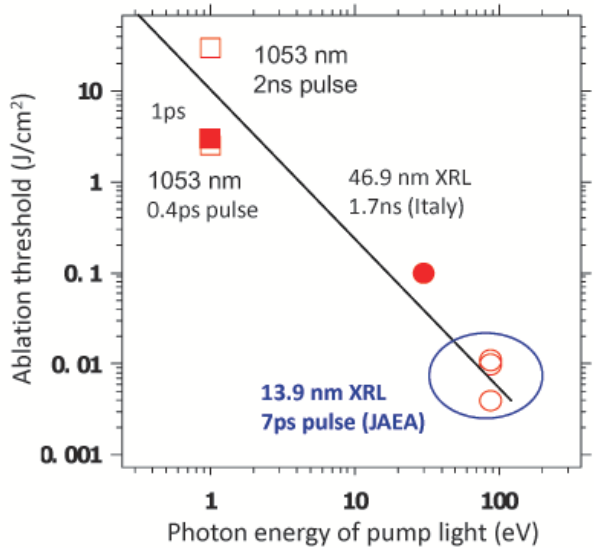}
   \caption{(Color on line)
   Comparison of the ablation thresholds (for incident fluence)
      as function of $\lambda_L$ and $\tau_L:$
        three squares \cite{diel-breakd-Rubenchik}, filled circle \cite{XRL-ns},
          two empty circles - this work. }
\label{fig:1}       % Give a unique label
\end{figure}

 % a-12

 We call a laser pulse short if $\tau_L<t_s.$
 Usually the condition $t_{eq}<t_s$ holds.
 In such a case,
   the rarefaction release,
     driven by $p_e,$
       envelopes smaller mass than the $p_i$-release,
   and nucleation takes place in one-temperature state.
 In the case when $t_{eq}>t_s$ expansion, subsequent stretching and nucleation
   are connected with $p_e.$
 In the two-temperature state,
   the vapor-solid and vapor-liquid coexistence curves
     are shifted in the direction toward the two-phase region.
 This is a result of a blow out by the electronic pressure of atomic system,
   composed of atoms which attract each other.
 At the same absorbed energy,
   the pressure $p_e$ is few times smaller
     than $p_i,$
       since the degenerate and classic gases of  electrons
         are softer
           and their Gruneisen parameters $\Gamma_e = 2/3$ is smaller
             than $\Gamma_i\approx 1.5-2.5$ for solids above the Debye temperature.

 % a-13 TermoMech 6 7 8,   9=ifsa

 The main condition for nanostructuring and spallative ablation is $\tau_L<t_s.$
 Other parameters,
   such as wavelength (from IR to X-ray)
     and spectra $(\Delta=0$ or $\Delta\neq 0),$
       are less significant.
 Therefore,
   spallative ablation is an important mechanism for removal of material.
 For IR and visible quanta $h\nu\sim 1$eV acting on metals and semiconductors,
   theory of spallative ablation is well developed \cite{LZreview,icpepa-2,Lorazo}.
 The theory is developed to the point
   where fine and secondary consequences of the metastable decay
     such as nanostructuring in case of a wide focal spot $d_{spot}\ll d_T$
         are predicted \cite{nanoreliefLETTjetp08,IFSA}.

 % a-14  cite=10 11   diel-Nr 12 13 14   15=Urb  16=Virginia

 For $h\nu\sim 1$eV and transparent dielectrics,
   the Newton rings \cite{1999,prl98} were never observed
     in spite of several attempts \cite{SKT-almaz,Siegel-germanium,Siegel-diel}.
 It seems that interplay of two circumstances makes this observation difficult.
 Indeed, significant absorption $E_{abs}$
   together with small absorption depth,
     similar to skin depth in metals,
       are achieved above threshold fluence $F_{brkd}$ for optical breakdown.
 At the same time,
   heating $E_{abs}$ is a very sharp function of absorbed fluence $F_{abs}$ near this threshold
     \cite{diel-breakd-vdLinde,diel-breakd-Rubenchik,diel-breakd-Quere,diel-breakd-Rethfeld}.
 Therefore,
   spallative ablation is limited to the narrow region near $F_{brkd},$
   because the value of heating is restricted $E_{abs}<E_{lim-up},$ $E_{lim-up}\sim 0.3 E_{coh}$
     \cite{Urbassek2008prb}.
 Above this limit,
   cohesive property is weak against strong stretching in a hydrodynamic rarefaction wave -
     expansion proceeds similar to expansion of heated gas (that is, without spallative plate).
 The cohesive property is responsible for creation of a spallative plate
   and spallative cupola \cite{virginia-1}.
 The cupola is necessary for the interference
   which results in appearance of the varying in time Newton rings \cite{virginia-1}.

 % a-15

 The second circumstance is connected with the width of the gap $\Delta.$
 The semiconductors such as Si and GaAs,
   with rather narrow $\Delta,$
     metallize during melting.
 In the meanwhile molten dielectrics remain in dielectric state.
 In this case,
    the cupola is dielectric, and Newton interference oscillations are weak
      (weak oscillations due to presence of oxide film
        have been detected in \cite{Siegel-germanium}).

 % a-16   17+18

 The Newton rings are bright manifestation of existence of spallative ablation.
 Appearance of rings means that a light wave interferes between spallative cupola and the rest of the target.
 This is a very surprising example of spallative plate so thin that it is even transparent (!)
   to light
     - the skin depth is of the order of 10 nm.
 In more customary spallation by a long laser pulse \cite{Eliezer,Krasyuk},
   the plate is much thicker.

 % a-17  XRL= 19 20 21 22 23

 Contrary to the case with quanta $h\nu\sim 1$eV and non-transparent substances
   like metals and semiconductors,
     the thermomechanical effects for X-ray irradiation are not investigated.
 This is the first attempt to study thermomechanics, metastability and nucleation induced by X-rays.
 This subject is interesting in connection with fast progress in developing of X-ray lasers
   and in connection with a whole number of experimental papers on the X-ray ablation
     \cite{KST-xrl,non-therm-xrl,xrl-met+sc+diel,FaenovInogamovAPL09,XRL-ns}.

 % a-18

 The paper is organized as follows:
   First, we show how the ratio $\tau_L/t_s$
     influences the maximum pressure created by absorption of laser energy.
 Then, new results concerning freezing of nanostructures at a late stage are presented.
 After that,
   the theoretical model of spallative X-ray ablation,
     and experimental findings connected with this model,
       are described.
 It is shown that as a result of the conditions,
   presented in the following lines (i), (ii) and (iii),
     the threshold $F_{abl}$ for the X-ray spallative ablation is extremely low,
       in comparison with other cases
         with different laser wavelength $\lambda_L$ and durations $\tau_L.$
 This is illustrated in Fig. \ref{fig:1}.
 The three mentioned conditions are:
   (i) 100\% absorption of X-rays (no reflection);
     (ii) negligible diffusion and heat conduction loses
       out from the small attenuation depth $d_{att};$
         and (iii) smaller energy densities necessary for spallative ablation,
           in comparison with evaporative ablation of equal amount of material.

\section{Duration of a laser pulse and amplitude of an acoustic response}
\label{sec:tauL-ts}

 Fast laser pulse transfers matter into energy containing state
   similar to a state of a chemical explosive behind a front of a detonation wave.
 This transfer is a base for hydrodynamic release, metastable decay and spallation.
 As was said, the pulse is short if its duration $\tau_L$
   is comparable or shorter than acoustic time $t_s.$
 In this case, a pathway at a thermodynamic phase plane consists of two parts.
 One is passed during a pulse $\tau_L,$ while the other - during an acoustic response $t_s.$
 The first part corresponds to the heating along the isochor $\rho=\rho_{initial}$
   with complications concerning the two-temperature details.
 The second part is formed by an approximately isentropic release along an adiabatic curve.
 It intersects the coexisting curve and penetrates into the two-phase region.

 % s-02 a-02

 The states of material near the target surface,
   irradiated by a long pulse $\tau_L\gg t_s,$
     are located near the coexistence curve.
 The hotter state relates to the larger absorbed laser intensity.
 Pressure created by a long pulse
   equals approximately to the pressure $p_{sat-vap}$ of saturated vapor.
 The later is limited by pressure $p_{cr}$ in the critical point,
   e.g., for Al $p_{cr} \sim 0.4$GPa.
 The critical pressure is significantly below pressures (\ref{eq:pi}), (\ref{eq:pe-2})
   achieved during spallative ablation.

 % s-02 a-03

 For the same final temperatures $T_{short}|_{bin}$ and $T_{long}|_{bin}$
   of material at the coexisting curve,
     in cases with short and long pulses,
       the maximum pressure $p_{short}|_{max}$ at the short pulse thermodynamic pathway
         is much higher;
           here the subscript "bin" marks the coexistence curve also called the binodal.
 A short pulse pathway
   deviates significantly from the coexistence curve
     into the high pressure condensed phase region.
 This is why the pressure $p_{short}|_{max}$ is higher.
 The $T_{short}|_{bin}$ corresponds to intersection of the part 2 of the short pulse pathway
   with the coexistence curve.
 The first paper, where the pathway has been used, was the paper \cite{prl98}.
 Now this useful conception is widespread \cite{nanoreliefLETTjetp08,LZreview,Lorazo,jetp99Barbel}.

 % s-02 a-04

 The temperature $T_{short}|_{bin}|_{abl}$
   corresponding to spallative ablation threshold
     is significantly below the critical temperature $T_{cr}.$
 Increase of the absorbed fluence $F_{abs}$ above $F_{abl},$
   increases $T_{short}|_{bin}$ above $T_{bin}|_{abl}.$
 There are distinct fluence $F_{ev}$ and temperature $T_{bin}|_{ev}$
   above which the spallative layer disappears \cite{virginia-1}.
 The ratio $T_{cr}/T_{bin}|_{ev}$
   depends on material properties and details of the two-temperature stage.
 It seems that it is larger for Au in comparison with Al.

 % s-02 a-05  MK=25

 For $F>F_{ev},$ material expands without spallation plate.
 This regime is different from the "phase explosion" \cite{MK} by long pulse
   with $T_{long}|_{bin}\approx T_{cr}.$
 Release of high pressure $p>0,$
   created isochorically by short pulse,
     produces expansion with high rate of stretching $\partial u/\partial x,$
       where $u$ and $x$ are along expansion direction.
 After nucleation in metastable state inside two-phase region
   the inertia of expanding matter inflates bubbles.
 The inertia is significant since the rate $\partial u/\partial x$ is high.
 Whereas in the case of the "phase explosion",
   the expansion of two-phase mixture
     is driven more slowly
       by weaker forces
         connected with pressure difference $\sim p_{sat-vap}-p_{out}$
           between pressure inside bubbles and pressure outside the target surface.

 % s-02 a-06

\begin{figure}
 \centering
  \includegraphics[width=1.0\columnwidth]{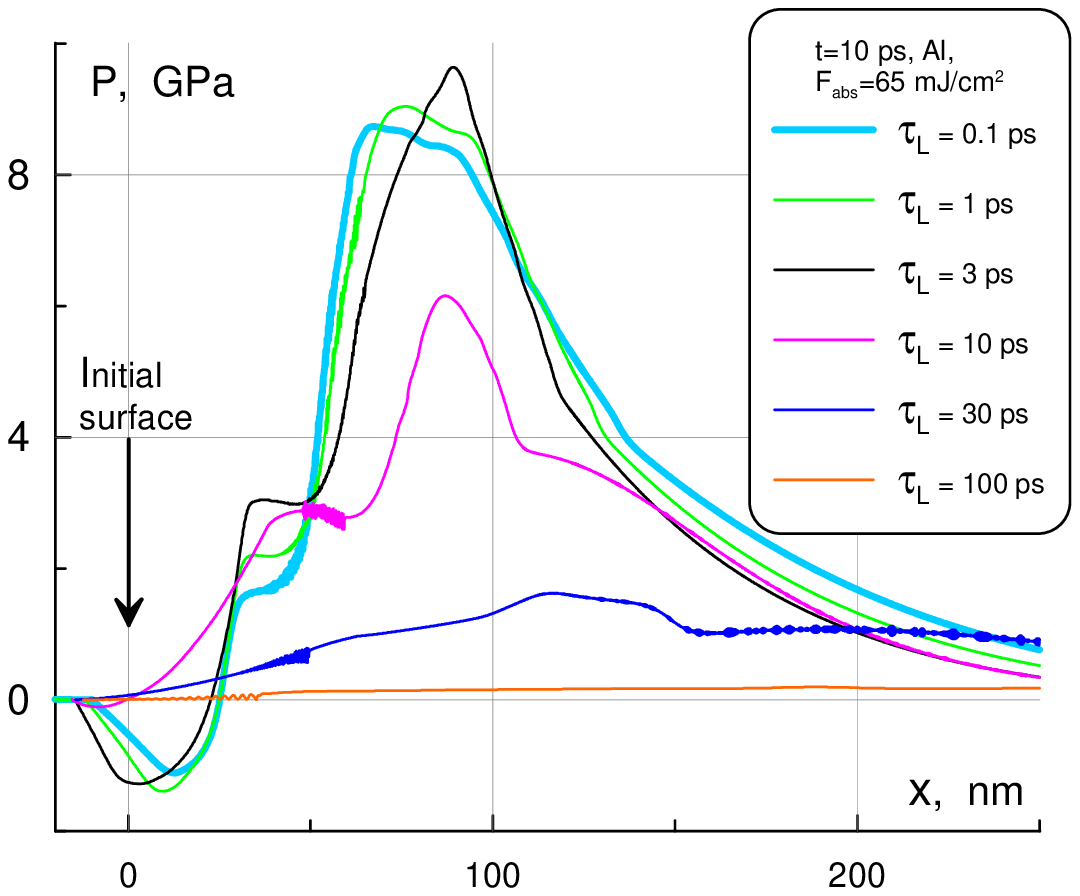}
   \caption{(Color on line)
    Strong decrease of maximum pressure for long pulses.}
\label{fig:2}       % Give a unique label
\end{figure}

 Importance of the ratio $\tau_L/t_s$ is illustrated in Fig. \ref{fig:2}.
 The acoustic time for this case is $t_s\approx 100\,{\rm nm}/(5.4\,{\rm km/s})=20$ps.
 We see that positive and negative pressures begin to decrease in their amplitude
   when the ratio $\tau_L/t_s$ becomes larger than unity.
 Distributions of pressure in the bulk Al target are shown at the instant $t=10$ps
   after the maximum of intensity $I\propto \exp(-t^2/\tau_L^2)$ of pump pulse.
 We use two-temperature hydrodynamic code described in Ref. \cite{icpepa-1}.
 Relaxation time $t_{eq}$ is of the order of 3 ps.

\section{Freezing of nanostructures}
\label{sec:Freezing-nanostr}

 % s-03 a-01

 As previously stated, a short pulse initiates a sequence of stages:
   (i) release, (ii) metastable state, (iii) nucleation, (iv) evolution of foam.
 Nucleation takes place if absorbed fluence is above nucleation threshold $F_{nucl}.$
 It is important that nucleation does not mean tacitly that spallative plate will run away.
 The spallative ablation threshold $F_{abl}$ and $F_{nucl}$ are separated.
 The separation $(F_{abl}-F_{nucl})/F_{abl}$ belongs to the few \% range.
 As was shown in \cite{nanoreliefLETTjetp08},
   during release,
     the foam inflates to thickness
   comparable to the distance between the foam and free surface of a target.
 Therefore it perturbs surface causing appearance of the surface nanorelief.

 % s-03 a-02       cite 27 jetp2008

 Development of the nanorelief is a dynamical phenomenon,
   caused by
     deceleration of free surface, inflation of foam, and surface tension resistance to inflation.
 There is the appearance time $t_{appr}$ when it develops.
 This time is much larger than $t_s$ because fluence $F$ is $\approx F_{abl},$
   and the nanorelief appears near the stopping point of free surface.
 Expansion velocities at this stage are small, and the nanoreleif develops slowly.
 Formation of the nanorelief may be experimentally observed as changes in reflectivity.
 There are changes due to surface nanorelief
   and due to absorption in thick foam under surface \cite{jetp2008}.

 % s-03 a-03

\begin{figure}[th]
 \centering
  \includegraphics[width=1.0\columnwidth]{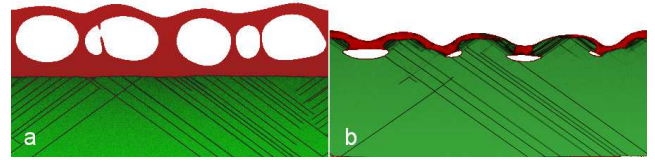}
\caption{(Color on line)
Freezing of Al foam as result of conductive cooling.
Green and red colors correspond to solid and liquid Al.
The melting/recrystallization front moves very slow at the time interval
 to which the left picture $t=154$ps belongs.
 At the right picture $t=468$ps the overcooled liquid layer with $T=570$K is partly frozen,
  the underlying bubbles are strongly deformed but survived.
 }
\label{fig:3-Al-nano-foam-evol}       % Give a unique label
\end{figure}

 In Al,
   and in many other materials,
     the ablation threshold $F_{abl}$
       is higher
         than the melting threshold $F_{melt},$
           and then at $F>F_{nucl}$ bubbles appears inside the molten layer.
 There are two cases with slow development of foam:
   one when $F_{nucl}<F_{abs}<F_{abl}$ and another when $F_{abs}$ is slightly above $F_{abl}.$
 In the first case,
   the foam remains closed under the surface.
 An example of this is shown in Fig. \ref{fig:3-Al-nano-foam-evol}.
 In the second case,
   there is process of slow detachment of the spallative plate.
 The plate is connected by the random net of the liquid filaments with foam.
 The filaments are stretched and break off one after another during detachment of the plate.
 During this process, a nanobrush from a forest of standing filaments appears.

 % s-03 a-04

 The subsequent evolution is material and target (bulk target versus foil, spot radial size) dependent.
 Let's comment first the dependence on material properties.
 In heavy metals late stages of foam development are especially slow.
 This is significant because there is competition
   between the late links of the foam development cascade
     and the rate of conductive cooling.
 E.g., density and material strength ratios for Au vs. Al are 19.3/2.7 and 20/13 \cite{icpepa-2},
   and gold foam moves slower.
 In Au, the layer between foam and melting front is thinner than in Al,
   freezing temperature is higher 1337/933,
     molten layer is thicker,
       and thermal conductivity 318/237 is higher;
         while heat capacity
           and surface temperature 2.5 kK at ablation threshold
             are approximately equal.

 % s-03 a-05     cite 28 = Vorobyev

 Cooling due to electron heat conduction is important.
 In liquid state,
   the foam under target surface $F<F_{abl}$ and nanobrush $F>F_{abl}$
     finally disappears
       because the surface tension collapses bubbles
         with low vapor pressure in the foam
           and smooths down the threads and surface bumps.
 Bubbles collapses when supporting them tensile stress decreases to zero.
 Contrary to this smoothing,
   the undersurface bubbles $F<F_{abl}$ and "nano-brush" $F>F_{abl}$ remain in the final relief
     if conductive loses are fast enough to freeze them.
 The nano-brush may be accompanied with frozen bubbles under.
 Comparison of Au vs. Al shows that in the case of Au, this is easier.
 This picture describes appearance of black gold in recent experiments
   \cite{nanoreliefLETTjetp08,Vorobyev}.
 Focusing of the laser beam and thickness of foil are also significant,
   since tight focusing improves cooling,
     while in thinner foils the reservoir to adopt heat from molten layer is smaller.

 % s-03 a-06

 Molecular dynamic (MD) simulation
   of development of foam and its freezing
     requires huge computer power and smart multi-processors algorithm,
       because foam occupies large volume
         and inflation/freezing processes are very slow.
 In the work \cite{nanoreliefLETTjetp08}
   the first part concerning the inflation of foam has been done.
 Here, we present new results describing freezing.

 % s-03 a-07

 In order to include the electron thermal conductivity in the MD code,
   the electrons as classical particles
     were added to atom subsystem.
 Each "electron"
   is assumed to associate with its host atom/ion.
 An electron always allocates the position of host,
   but it has its own velocity.
 The neighboring atoms may exchange their electrons
   with some frequency,
     determined from known experimental thermal conductivity
       of liquid aluminum/gold at melting point.
 The pair of atoms,
   where electron-electron exchange takes place,
     is chosen randomly from
        the list of neighbors of each atom.
 The list contains neighboring atoms
   within cutoff distance.

 % s-03 a-08

 In addition to the frequency of electron-electron exchange between atoms,
   we adjusted the mass of "electrons",
     in order to obtain the characteristic time
       of electron-ion energy exchange in collisions.
 The electron-ion scattering
   depends on the velocity vectors of an electron with mass $m_1$
     as a projectile and target ion with mass $m_2$
       according to equations:
 \begin{equation}
 \mathbf{v}_1^\prime=
 (m_1\mathbf{v}_1 + m_2\mathbf{v}_2 + m_2 v\mathbf{n})/(m_1+m_2)  \label{eq:e-e_02}
 \end{equation}
 \begin{equation}
 \mathbf{v}_2^\prime=
 (m_1\mathbf{v}_1 + m_2\mathbf{v}_2 - m_1 v\mathbf{n})/(m_1+m_2)  \label{eq:e-e_03}
 \end{equation}
 where $v$ is an electron-ion relative speed,
   and $\mathbf{v}_1^\prime$ and $\mathbf{v}_2^\prime$
     are the velocities of particles in the laboratory system after collision.
 The unknown unit vector $\mathbf{n}$
   is assumed to be distributed isotropically
     and is produced by using a generator of random directions.
 The Eqs. (\ref{eq:e-e_02},\ref{eq:e-e_03})
   conserve the momentum and total energy of electron and ion subsystems.
 Moreover,
   such a combined Monte-Carlo-MD approach
     guarantees conservation of local charge neutrality.

 % s-03 a-09

 Our tests indicate,
   that the described above electron-electron exchange
     and electron-ion collision procedures
       lead to energy diffusion through dynamic net of atoms,
         simulated by MD,
           and give a correct solution
             of continuum heat conduction equation for liquid aluminum/gold.

 % s-03 a-10  cite 29 Duff,   30 Chan=Ag-melt

 Two stages of evolution of Al nanobubble chain
   are shown in Fig. \ref{fig:3-Al-nano-foam-evol}.
 Fluence in this case is slightly below spallative ablation threshold $F_{abl}.$
 Therefore spallative plate keeps its connection to target
   after the process of inflation of bubbles and freezing.
 Our approach differs from the approach applied in the important paper \cite{Duff},
   where the one-dimensional (1D) finite-difference scheme has been used.
 The scheme takes into account 1D thermal conductivity
   and is solved parallel to MD simulation.
 The 1D scheme employs transversally averaged temperature fields
   from the parallel MD simulation.

 % s-03 a-10.2

 It is important that in case of foam
   our method allows to describe spatial separation of thermal fluxes
     pumping heat through threads and walls with small cross-section.
 This greatly delays cooling through foam.
 Therefore the base of the foam,
   which is located from the bulk side,
     cools faster
       than the film between the foam and free surface
         as it is shown in Fig. \ref{fig:3-Al-nano-foam-evol} (b).
A liquid layer above the bubbles in Fig.~\ref{fig:3-Al-nano-foam-evol} (a)
   has temperature gradient from $1380$K on its free surface to $1340$K at its internal surface
   contacting with bubbles.
The liquid layer shown in Fig. \ref{fig:3-Al-nano-foam-evol} (b) is in supercooled state,
   comp. with \cite{Ag-melt}.
 Its temperature is $T\approx 570$K which is below melting point $933.6$K on 360 K,
and temperature gradient across the layer is small in this case.
 Separated nanocrystallites start to grow from two free surfaces of the liquid layer.
As \ref{fig:3-Al-nano-foam-evol} indicates a bubble position is not random but correlated with positions of
second and third neighbors at least. The freezing process fixes this correlation with formation of surface
nanostructures having correlation length a few inter-bubble distance.

%    i pro dolgo nezastivayuchshuyu plenku **** *****

% (a) map of symmetry factor $s$ at t = ? ps, green corresponds to soild while red - to liquid,
% In the case considered there are nucl, foam formation, then inflation of bubbles in foam.
% during the inflation
%   the foam degrades
%     as bubbles become large and number of bubbles per target unit area decreases.
% the stage when ostaetsya po odnomu puziryu po glubine.  NET,ETO Al OKOLO POROGA  TAM TSEPOCHKA PUZIREI=
% gde front plavleniya.
% eto primerno na moment max udaleniya fronta plavleniya --prover'.
% dalee nachinaetsya kristallizatsiya.
% front kristlz dvizhetsya k svob gran (ona sverkhu) - tolchsh rasplava umen'shaetsya.
% nachinaetsya zamerzanie zhidkosti vokrug puzirei so storoni bulk.
% pri etom puziri kontragiruyut=szhimayutsya - prichem v osnovnom v normal'nom k poverkhnosti napravlenii.
% dolgoe zamerzanie plenki.
% (b) t=??500ps a ona echshe ne zamerzla.
% esli b ne teploizolyatsii plenki puziryami - to rasplav davno zamerz bi
%  (kogda bi on zamerz?prover' po kinogramme)

 % s-03 a-11 cite 31=movies at the ITF site

 The foam may be partially frozen from the bulk side,
   when, in the case with fluence above ablation threshold $F_{abs}>F_{abl},$
     still liquid film between the foam and free surface of target
       begins to transform to the spallative plate
         and the process of breaking of the liquid-solid threads starts.
 In this case the forest of the frozen threads appears
   (black gold \cite{nanoreliefLETTjetp08,Vorobyev}).
 MD simulation shows that the gold foam has a thick multi-level structure
   with significant drop of temperature and expansion rate $\partial u/\partial x$
     through levels from one level to another
       toward the bulk side, see movies \cite{DepLaser}.
 Detailed description of the MD simulation results,
   concerning foaming and freezing,
     is beyond the scope of our paper.
 Here we have to show that a short X-ray laser pulse may cause similar thermomechanical phenomena
   as an optical pulse.
 Thick foam is formed if the fluence $F$ is significantly above the threshold $F_{abl}.$
 The "flashes" of the successive nucleations during a process of formation of thick foam
   take place
     when positive pressure $p$ is lowered by the decrement $p_{str}$ \cite{jetp2008},
       where $p_{str}$ is material strength.
 At acoustic stage pressure is halved from the initial value to the value
   corresponding to the established compression wave \cite{icpepa-2}.
 If $p_{str}$ does not depend on temperature then the distances between successive flashes are equal.
 In case $p_{str}(T)$ the distances is growing toward the bulk
   since temperature decreases and $p_{str}(T)$ increases.

 % s-03 a-12

 Near threshold $F_{abl}$ the foam evolution time $\sim 0.1-1$ns
   greatly exceeds collisional relaxation times $\sim t_{eq}\sim 1-10$ps
     for $h\nu\sim 1$eV and X-ray lasers.
 Therefore the picture with melting and foam
   may be applied to the case of dielectrics heated by short pulse X-ray laser.
 Similar phenomena may appear in case of metals and semiconductors.
 In the case of LiF considered below
   the ablation threshold $F_{abl}$ is lower than the melting threshold $F_{melt}.$
 The work \cite{Urbassek2008prb} is devoted to comparison of the spallative ablations
   from liquid and solid states.
 If we rise fluence up to $F_{melt}$ in LiF
   then deceleration of free surface to almost stopping point will be not possible,
     because to slow down motion it is necessary to fulfill the condition $F\approx F_{abl}.$
 This means that in LiF near $F_{abl}$ there are no melting/freezing processes described above.
 But other substances, e.g. Al, have $F_{melt}<F_{abl}.$
 In these cases X-ray laser will cause nanostructuring.

\section{Theoretical model of X-ray action}
\label{sec:XRL-theor}

 % s-04 a-01

\begin{figure}
 \centering
  \includegraphics[width=1.0\columnwidth]{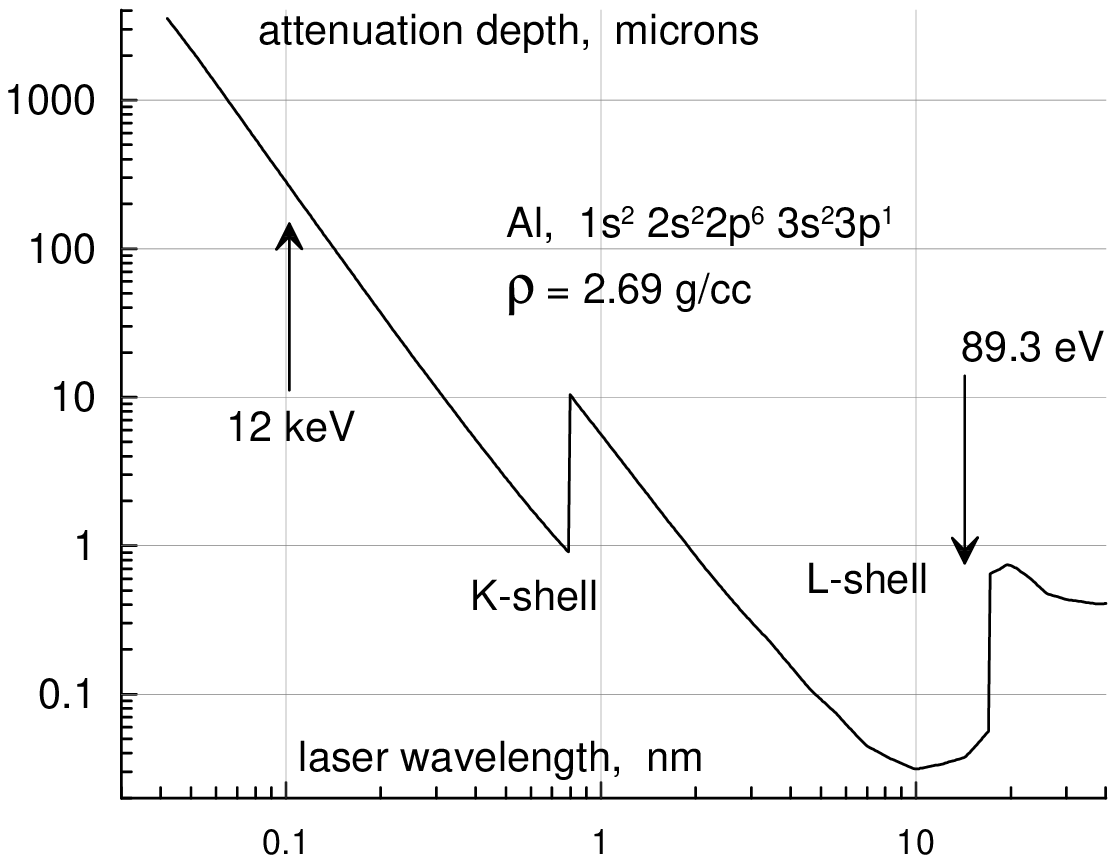}
   \caption{Variation of $d_{att}(h\nu)$ with photon energy $h\nu.$}
\label{fig:04-dAtt}       % Give a unique label
\end{figure}

 Dependence of the attenuation depth $d_{att}(h\nu)$ taken from \cite{Henke}
   is shown in Fig. \ref{fig:04-dAtt}.
 The right arrow corresponds to the Ag X-ray laser described in next Section.
% At high $h\nu$ dependencies $d_{att}(h\nu)$ are similar for different substances.
 The depth $d_{att}$ is large for high energy photons.
 Then the acoustic response $t_s=d_{att}/c_s$ corresponding to hard photons is long.
 For energy $\hbar\omega_L=12\un{keV},$
   which will be achieved soon at XFEL \cite{XFEL},
     it is $t_s \sim 50$ns.
 This is very large time in comparison with the picosecond time scale of the atomic processes.

 % s-04 a-02

 The estimate of the spallative ablation threshold for these hard photons is
   $$
   F_{abl}=\zeta \, d_{att}\, n_{at}\,E_{coh} \approx 200\un{J/cm}^2,
   $$
     where $\zeta=0.2-0.4$ is a coefficient \cite{icpepa-2,Urbassek2008prb}
       in relation $E_{abl}=\zeta\,E_{coh},$
         $E_{abl}$ is energy per atom at the threshold,
           $n_{at}$ is the atom concentration in solid state.
 Future XFEL lasers \cite{XFEL} will have $0.1\un{keV}<\hbar\omega_L<12\un{keV}$ photons,
   ultrashort durations $\tau_L\sim 20\un{fs},$
     and fluence up to several hundred J/cm$^2.$
 Irradiation by such pulse can cause spallative ablation of huge sub-millimeter piece of condensed target.

 % s-04 a-03

 Equations describing the electron-ion non-equilibrium stage and hydrodynamic motion are
\begin{equation}
\partial x/\partial t=u,\; \rho\partial x=\rho^o\partial x^o,\,
  \rho^o\partial u/\partial t = - \partial p/\partial x^o,
          \label{eq:kinem-eqs-1+2+dyn-3}
\end{equation}
\begin{equation}
 \rho^o\frac{\partial E^{sum}_e/\rho}{\partial t} = - \frac{\partial q_e}{\partial x^o}
  - p_e\frac{\partial u}{\partial x^o}-\frac{\rho^o}{\rho}\dot E_{ei}+\frac{\rho^o}{\rho}Q,
          \label{eq:Ee-4}
\end{equation}
\begin{equation}
 \rho^o\frac{\partial E_i/\rho}{\partial t} =  - \frac{\partial q_i}{\partial x^o}
  - p_i\frac{\partial u}{\partial x^o} + \frac{\rho^o}{\rho}\dot E_{ei},
          \label{eq:Ei-5}
\end{equation}
\begin{equation}
 \rho^o\frac{n_e/\rho}{\partial t} = - \frac{\partial j}{\partial x^o}
   + \frac{Q}{u_{i2}} + \nu_{imp} n_e - \kappa_{rec} n_e^3
          \label{eq:ne-6}
\end{equation}
 This is system of equations in Lagrangian coordinate $x^o$
   corresponding to an initial position of a material point.
 The initial density profile inside the target is $\rho(x,t = - \infty) = \rho^o.$
 Values $x,$ $u,$ $p=p_e+p_i,$ $E^{sum}_e=n_e u_{i2}+E_e,$ $E_i,$ $n_e$
   in (\ref{eq:kinem-eqs-1+2+dyn-3}-\ref{eq:ne-6})
     are functions on variables $x^o,t;$
       $n_e$ is electron concentration $n_e$ in the conduction band;
         $\dot E_{ei}=\alpha\,(T_e-T_i)$ is an electron-ion energy exchange rate;
           $q_{e,i} = - (\rho\kappa_{e,i}/\rho^o) \partial T_{e,i}/\partial x^o$ are heat fluxes;
             $j = - (\rho\, D/\rho^o) \partial n_e/\partial x^o$ is diffusion flux of electrons;
               $\kappa_{e,i}$ are electron and ion thermal conductivities.

 % s-04 a-04

 Primary electrons and primary holes are produced during X-ray pulse acting on LiF.
 Their kinetic and potential energies are
   $E_e|_{prim}\sim 10-30$eV and $u_{i1},$ $E_e|_{prim}+u_{i1}=h\nu=89.3$eV.
 These energies are measured from the bottom of the conduction band.
 Primary electrons and holes are non-equilibrium ones -
   they are described by the distribution function -
     they can not be described by temperature $T_e$ and concentration $n_e.$
 The non-equilibrium electrons and holes relax to thermalized state with secondary electrons and holes
   through Auger processes, impact ionization, and three-body recombination.
 Relaxation time is $\tau_{rel}\sim 1$ps.
 The value $\tau_{rel}$ is smaller than the pulse duration $\tau_L=7$ps.
 Therefore,
   if we exclude the short time lapse $\tau_{rel}$ in the beginning of the laser pulse,
     then we can include contribution of primary particles into electron energy budget (\ref{eq:Ee-4})
       through energy conservation (radiative loses are small).
 It is supposed in equations (\ref{eq:Ee-4}) and (\ref{eq:ne-6})
   that the laser source $(\rho^o/\rho) Q$ supplies energy directly to secondary electrons
     and produces secondary electrons at the rate $Q/u_{i2},$
       where $u_{i2}\approx \Delta$ is an ionization potential of secondary electrons,
         $\Delta\approx 14$eV is width of forbidden gap in LiF.

 % s-04 a-05

 Every primary electron produces $(h\nu=89.3\un{eV})/(u_{i2}+E_e)\approx 6$ secondary electrons.
 Solution of system (\ref{eq:kinem-eqs-1+2+dyn-3}-\ref{eq:ne-6}) for parameters
   $\tau_L=7$ps, $F=10$mJ/cm$^2,$ LiF, $d_{att}=28$nm
      gives $n_e|_{max}=(1-2\%)n_{at},$ $T_e|_{max}\sim 2$eV,
         where $|_{max}$ corresponds to the maximum values.
 These values are achieved at the end of X-ray laser pulse.
 The parameters correspond to the experiment described in next Section.
 Electrons are classical since their kinetic energy $(3/2)k_B T_e\sim 3$eV is larger
   than Fermi energy $E_F\sim 0.5$eV
     corresponding to our case with low concentration of free electrons.

 % s-04 a-06

\begin{figure}
 \centering
  \includegraphics[width=1.0\columnwidth]{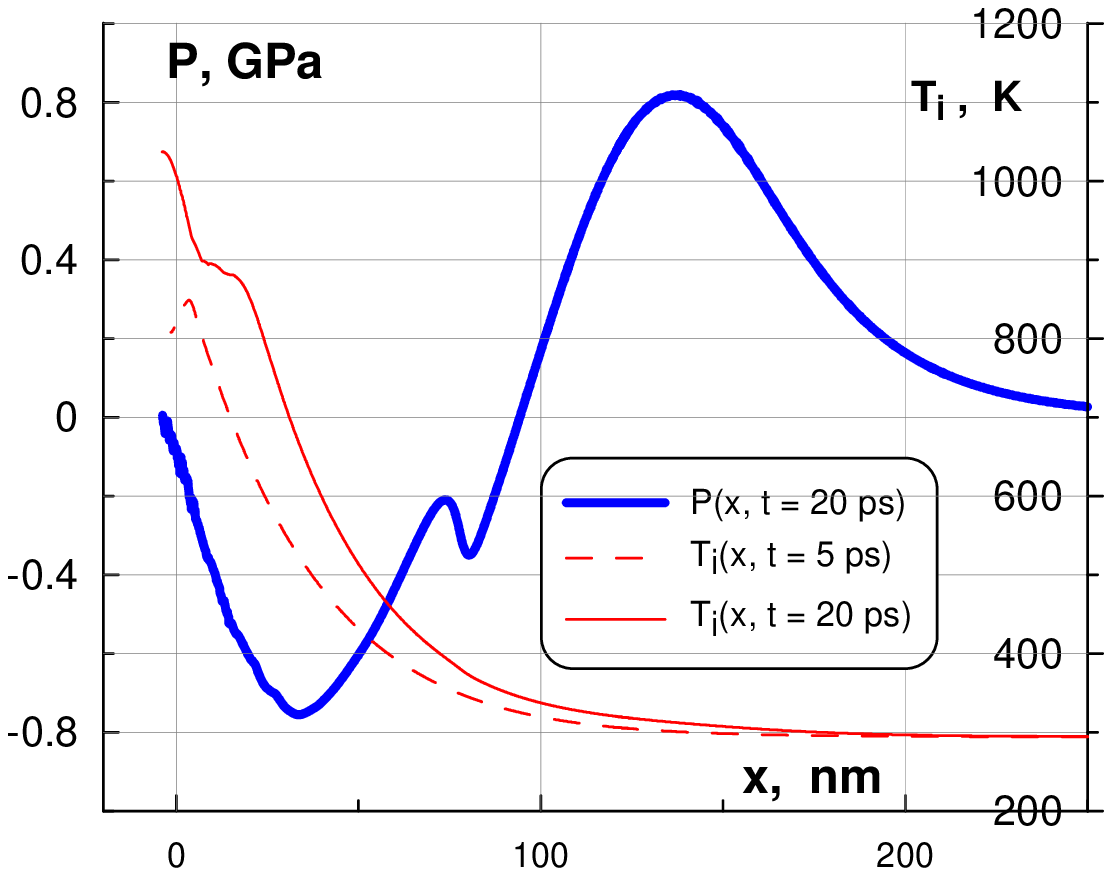}
\caption{(Color on line)
 Pressure and temperature profiles in LiF irradiated by X-ray laser pulse.
 }
\label{fig:5}       % Give a unique label
\end{figure}

 The energy transfer rate is calculated as
   $\dot E_{ei}=\alpha\,(T_e-T_i)\approx$ $\alpha T_e = A E_e,$ $A=2\alpha/3k_B,$
     since heating of a lattice by electrons
       is significant only when electrons are much hotter than a lattice.
 Lattice temperature for $F=10$mJ/cm$^2$ is $T_{at}|_{max}\approx 700$K.
 Pressure wave initiated by this fast increase of temperature is shown in Fig. \ref{fig:5}.
 The wave travels to the bulk side of a target (to the right side in Figure) from the irradiated surface,
   while temperature profile remains "frozen" into matter.
 The point $x=0$ is an initial position of the surface.
 The profile of the wave contains tail with negative pressure $p_{neg}$ (tensile stress).
 The amplitude $|p_{neg}(t)|$ is enough to cause spallative ablation.
 Crater depth is 40-50 nm.
 These findings agree with experimental results described below.
 % increases gradually with time
 %  and saturates at the amplitude $p_{sat}$ at the point in time $t_{sat}\sim t_s=d_{att}/c_s.$
 %  After that tensile stress does not change.

% теор.модель =сылка на спп. а что дальше?? развитие дл€ —ј+студент мифи облуч √ајр с обратн стороны

\section{X-ray experiment}
\label{sec:XRL-expr}

  % s-05 a-01

\begin{figure*}
\vskip 5mm
 \centering
  \includegraphics[width=0.75\textwidth]{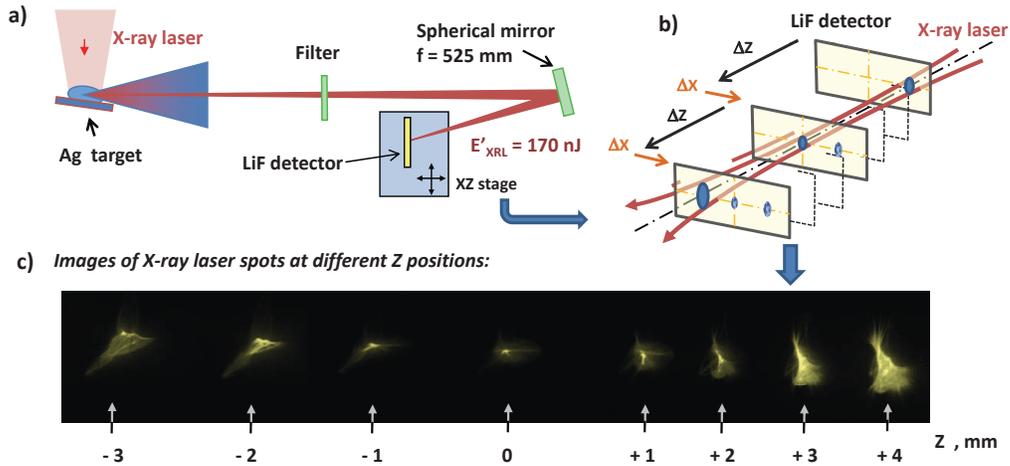}
   \caption{\label{fig:PNP_1} (Color online)
 (a) The experimental set up for recording the Ag XRL beam patterns near the best focus position on a LiF crystal.
 (b) Sketch of motion of a LiF crystal during experiments.
 (c) The patterns of XRL beam focusing spots recorded on a LiF crystal
     at -3 mm to +4 mm from the best focus position. }
\end{figure*}

 The experiment has been performed with the Ne-like Ag soft x-ray laser (XRL) facility
   at JAEA Kansai Photon Science Institute, working at transient collisional scheme
     \cite{FaenovInogamovAPL09,Nishikino,Faenov09}.
 The XRL beam with an energy $\sim 1$microJ
   and the horizontal and vertical divergences of 12 mrad$\times$5 mrad, respectively,
     was focused on a LiF crystal of 2 mm thickness and 20 mm diameter,
       by using a spherical Mo/Si multilayer mirror of 1050 mm radius of curvature
         (see Fig. \ref{fig:PNP_1}).
 The total energy on the LiF crystal of the XRL beam after passing 200 nm Zr filter
   and reflecting from the focusing mirror was $\sim 170$nanoJ in a single shot.
 The luminescence of stable color centers (CCs)
   \cite{4-F,5-F,6-F,7-F}
      formed by XRL radiation,
        was used to measure the intensity distribution in the XRL laser focal spot
          \cite{FaenovInogamovAPL09,Faenov09}.
 After irradiation of the LiF crystal with the XRL,
   the photo-luminescence patterns from the color centers (CCs) in LiF (shown in Fig. \ref{fig:PNP_1})
     were observed by using a confocal fluorescence laser microscope (OLYMPUS model FV300).
 An OLYMPUS BX60 microscope in visible differential mode
   and an atomic force microscope (AFM, TOPOMETRIX Explorer), operated in the tapping mode,
     have been used for measurements of the size of ablative spot.
 As it was shown in our previous experiments \cite{Faenov09},
   only about 6\% of full laser energy
     is concentrated to the best focus spot of $\sim 200$square microns.
 This corresponds to energy $\sim 5$mJ/cm$^2$ or laser intensity $\sim 7\times 10^8$W/cm$^2.$

% For two-column wide figures use

   % s-05 a-02

\begin{figure*}
\vskip 8mm
 \centering
  \includegraphics[width=0.65\textwidth]{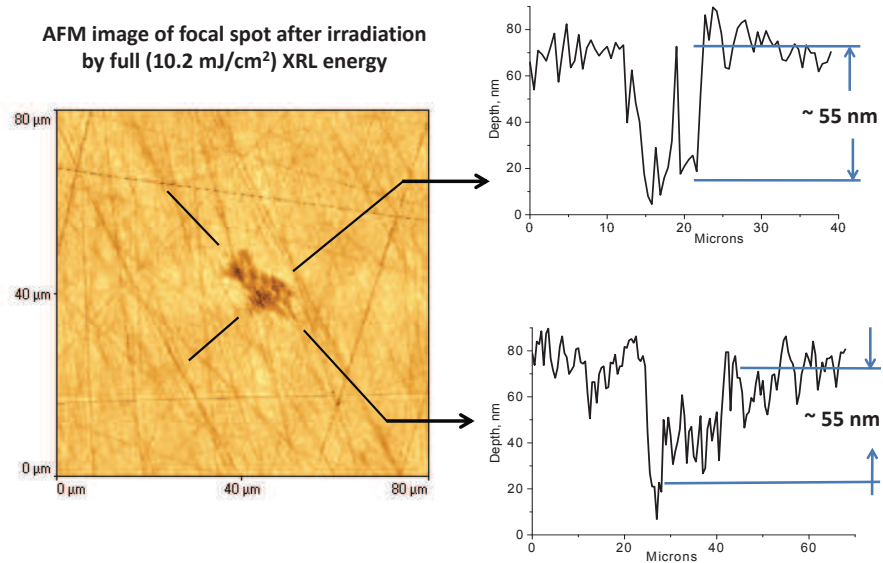}
   \caption{\label{fig:PNP_2} (Color on line)
 AFM image and traces, taken through orthogonal directions, of the ablative spot on LiF crystal,
 irradiated by single shot with full (10.2 mJ/cm2) laser intensity of XRL beam pulse.  }
\end{figure*}

 Two types of experimental investigations of XRL ablation threshold of LiF crystals were done.
 In the first experiments the Zr filter has been removed,
   and the XRL beam expands without attenuation by the filter.
 This powerful beam has been focused on the surface of LiF crystal.
 In Fig. \ref{fig:PNP_2} the AFM image of the focal spot of this XRL beam is presented.
 The image is obtained after a single laser shot.
 Ablation of crystal is clearly seen at the AFM image and traces.
 The ablation threshold for LiF
   irradiated by a single shot
     is $10.2$mJ/cm$^2.$
 From the traces in Fig. \ref{fig:PNP_2},
   we could see that the ablation depths varied between 30 and 55 nm.
 These values are close to the theoretical crater depths calculated above.

   % s-05 a-03

\begin{figure*}
 \centering
  \includegraphics[width=0.75\textwidth]{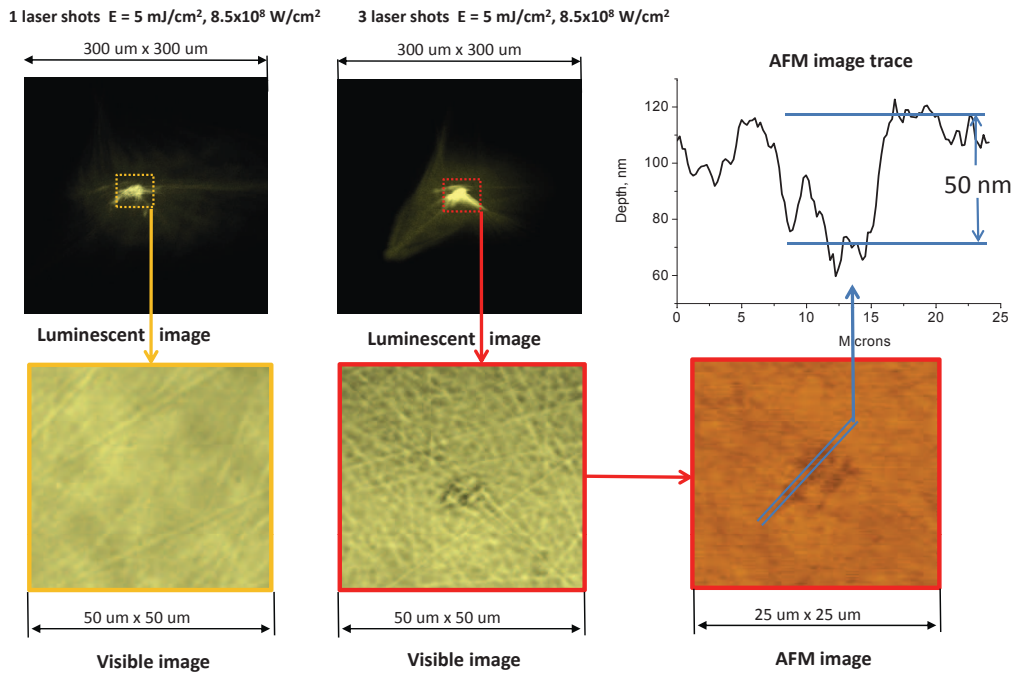}
   \caption{\label{fig:PNP_3} (Color on line)
 (a) The luminescence, visible and AFM images of XRL beam focusing spots on the surface of a LiF crystal
    obtained after one and three shots with XRL laser intensity 5 mJ/cm$^2.$
 Trace was done for AFM image, obtained in the case of three shot irradiation of a LiF crystal.}
\end{figure*}

 In the second type of experiments,
   the Zr filter was settled inside the propagation path of the XRL beam.
 Three shots have been done in the same focusing spot
   with fluence of 5 mJ/cm$^2$ for each shot.
 We could see in Fig. \ref{fig:PNP_3} that in this case,
   a crater,
     with an ablation depth of about 50 nm,
       appears on the surface of the crystal.
 This is similar to the first experiment with single more intensive XRL shot.

    % s-05 a-04

 The ablation threshold
   obtained in our experiments
     is much smaller in comparison with previous experiments, see Fig. \ref{fig:1}.
 Our threshold is 3400, 300, and 10 times smaller
   than thresholds for nanosecond and femtosecond Ti:sapphire lasers,
     and for nanosecond 46.9 nm soft XRL, respectively.

\section{Conclusion}
\label{sec:1}

 It is shown that
   short pulse of XRL causes thermomechanical response
     as in cases with optical lasers.
 There is spallative ablation as result of appearance of tensile stress
   which overcomes material strength above ablation threshold.
 Near threshold stretching of melt may be accompanied by nanostructuring.
 Nanostructures freeze down if conductive cooling is fast.
 Value of the X-ray threshold is small in comparison with irradiation by longer wavelengths
   and/or longer pulse.

 Work has been supported by the RFBR grant No. 09-08-00969-a (NAI, VVZh, VAK, and YuVP).
 This research has been partially supported by the Japan Ministry of Education, Science, Sports and Culture,
   Grant-in-Aid for Kiban A No 20244065, Kiban B No. 21360364
     and by the RFBR grant No. 09-02-92482-MNKS-a (AYaF, IYuS, TAP).

%
% BibTeX users please use
% \bibliographystyle{}
% \bibliography{}
%
% Non-BibTeX users please use

\end{document}